\documentclass[%
 aip,
 amsmath,amssymb,
 reprint,%
superscriptaddress
]{revtex4-1}

\usepackage{graphicx}
\usepackage{dcolumn}
\usepackage{bm}

\usepackage[utf8]{inputenc}
\usepackage[T1]{fontenc}
\usepackage{mathptmx}
\usepackage{etoolbox}

\makeatletter
\def\@email#1#2{%
 \endgroup
 \patchcmd{\titleblock@produce}
  {\frontmatter@RRAPformat}
  {\frontmatter@RRAPformat{\produce@RRAP{*#1\href{mailto:#2}{#2}}}\frontmatter@RRAPformat}
  {}{}
}%
\makeatother
\begin{document}

\preprint{AIP/123-QED}

\title[Multiscale super-resolution flow reconstruction]{Multiscale super-resolution reconstruction of fluid flows with deep neural networks}
\author{G.C. Yang}
 \email{yanggch8@mail.sysu.edu.cn}
\author{R.Y. Luo}%
\author{Q.H. Yao}%
\affiliation{School of Aeronautics and Astronautics, Sun Yat-sen University, Guangzhou 510275, China}

\author{P.J. Wang}
\affiliation{%
Institute of Intelligent Ocean Engineering, Harbin Institute of Technology (Shenzhen), Shenzhen, 518055, China
}%

\author{J.X. Zhang}
\affiliation{School of Aeronautics and Astronautics, Sun Yat-sen University, Guangzhou 510275, China}

\date{\today}

\begin{abstract}
We present a novel multiscale super‐resolution framework (SRLBM) that applies deep learning directly to the mesoscopic density distribution functions of the lattice Boltzmann method for high‐fidelity flow reconstruction. Two neural network architectures, a standard convolutional neural network (CNN) and a deeper residual dense network (RDN), are trained to upscale distribution functions from coarse grids by factors of 2, 4 and 8, and then recover velocity, pressure, and vorticity from a single model. For flow past a single cylinder at $\mathrm{Re}=100$, RDN reduces the mean relative error in distribution functions by an order of magnitude compared to CNN and avoids spurious pressure oscillations and vorticity smoothing that affect interpolation and simpler networks. To examine the generalization ability, both models are trained using data from the flow past two cylinders of diameter $d$ at a spanwise distance between the centers of $1.5d$ and a Reynolds number of 200. They are then applied without retraining to wake configurations with distances ranging from $2.0d$ to $3.0d$. In these tests, the mean errors remain essentially unchanged across all distances. However, RDN consistently produces sharper shear-layer roll-ups and secondary eddies. These results demonstrate that super‐resolving mesoscopic distribution functions yields richer and more transferable features than operating on macroscopic fields alone. By integrating kinetic theory with deep learning, SRLBM offers a compelling alternative for fluid flow reconstruction, enabling a single model to simultaneously recover multiple high-fidelity flow fields while substantially reducing computational cost.
\end{abstract}

\maketitle


\section{\label{sec:intro}Introduction}

High‐resolution flow fields are indispensable for applications ranging from aerodynamic optimization of aircraft, automobiles, and wind turbines to prediction of fluid transport in human airways and blood vessels. Achieving such resolution in computational fluid dynamics (CFD) requires extremely fine spatial meshes and small time steps to capture the smallest flow scales, yet this requirement becomes computationally prohibitive at high Reynolds numbers \citep{choiGridpointRequirementsLarge2012}. At the same time, preliminary design studies and closed‐loop feedback control systems frequently require rapid CFD predictions with acceptable accuracy \citep{vinuesaEmergingTrendsMachine2022}. Consequently, when computational resources are limited, reconciling high‐fidelity simulation with practical efficiency remains a fundamental challenge in fluid flow modeling.

To alleviate the prohibitive computational and memory demands of direct numerical simulation (DNS), researchers have developed more efficient paradigms, including large‐eddy simulation (LES), Reynolds‐averaged Navier-Stokes (RANS) modeling, and reduced‐order modeling (ROM). LES reduces cost by applying a spatial filter that removes motions below a prescribed cutoff and directly resolving only the large, energy‐containing eddies on coarser grids and larger time steps. The filtered subgrid-scale stresses are then modeled by turbulence closures \citep{smagorinskyGeneralCirculationExperiments1963, lesieurNewTrendsLargeeddy1996}. RANS achieves even greater efficiency by replacing instantaneous equations with time‐ or ensemble‐averaged forms, thus solving for steady mean fields while modeling all turbulent fluctuations through closure relations \citep{chenSolutionsReynoldsaveragedNavierStokes1990, alfonsiReynoldsAveragedNavierStokes2009}. As a result, LES typically delivers fidelity higher than that of RANS, since it computes the dominant eddies responsible for most of the momentum and scalar transport. ROM attains the greatest speed by projecting the flow dynamics onto a low‐dimensional basis of coherent structures, often obtained by proper orthogonal decomposition or dynamic mode decomposition, exploiting the fact that many complex flows are governed by a few dominant modes \citep{rowleyModelReductionFlow2017, tairaModalAnalysisFluid2017, tairaModalAnalysisFluid2020}. However, ROMs often depend on bases tailored to specific flow configurations, which constrains their robustness and generality across different geometries and operating conditions \citep{vinuesaEnhancingComputationalFluid2022}.

In recent years, machine learning has emerged as a transformative paradigm for enhancing both the accuracy and efficiency of traditional CFD methods. \citet{stevensEnhancementShockcapturingMethods2020} trained a neural network to correct truncation errors in a fifth-order shock-capturing scheme, markedly sharpening its resolution of discontinuities. \citet{ajuriaillarramendiHybridComputationalStrategy2020} then introduced a hybrid CFD-CNN framework that learns an approximate pressure Poisson solver, thereby reducing the iterative cost of incompressible flow simulations. \citet{jiangNeuralNetworkBasedPoisson2024} built on this idea by combining a residual neural network with an iterative correction loop, achieving an order‐of‐magnitude speedup in solving the Poisson's equation without compromising accuracy. In the Reynolds‐averaged Navier-Stokes context, \citet{lingReynoldsAveragedTurbulence2016} embedded Galilean invariance into a neural network architecture to predict the full anisotropic Reynolds stress tensor, outperforming classical linear and nonlinear eddy viscosity models. Data-driven subgrid scale closures for LES have been realized by training deep networks to map coarse-resolved flow fields to closure terms \citep{gamaharaSearchingTurbulenceModels2017, beckDeepNeuralNetworks2019, maulikSubgridModellingTwodimensional2019}. In reduced order modeling, autoencoder networks compress high-dimensional flow fields into low‐dimensional latent spaces, where subsequent dynamics are predicted via purely data-driven or hybrid physics‐informed models, yielding substantial computational savings \citep{murataNonlinearModeDecomposition2020, leeModelReductionDynamical2020, vlachasDatadrivenForecastingHighdimensional2018, pathakModelFreePredictionLarge2018, callahamRoleNonlinearCorrelations2022}. Similar techniques have recently been applied to the lattice Boltzmann method, where learned compression schemes and data‐driven closure corrections promote both speedups and fidelity enhancements \citep{hennigh2017latnetcompressinglatticeboltzmann, chenCompressedLatticeBoltzmann2021, corbettaLearningLatticeBoltzmann2023}. Collectively, these advances illustrate how machine learning can integrate data‐driven insights with governing equations to bridge the fidelity-efficiency gap in fluid-flow simulation.

Machine learning based flow reconstruction has emerged as a powerful alternative for recovering high-fidelity fields from sparse or coarse data. Inspired by breakthroughs in computer vision, \citet{fukamiSuperresolutionReconstructionTurbulent2019, fukamiSuperresolutionAnalysisMachine2023} pioneered the application of neural networks to super‐resolution (SR) turbulent flow fields, demonstrating that learned SR models outmatch bicubic interpolation by accurately recovering high-wavenumber structures without excessive smoothing. Subsequently, \citet{liuDeepLearningMethods2020} introduced a multi-temporal-path CNN that leverages consecutive snapshots to better capture the unsteady eddies of turbulence, and \citet{xuSuperresolutionReconstructionTurbulent2023} showed that transformer architectures can simultaneously learn isotropic and anisotropic features in complex geometries. To circumvent the need for high-resolution (HR) ground truth, \citet{gaoSuperresolutionDenoisingFluid2021} embedded conservation laws and boundary conditions directly into a CNN loss function, yielding a physics-informed SR approach that enforces fluid dynamics constraints a priori. More recently, \citet{chuFlowReconstructionSUBOFF2024a} leveraged physics-informed neural networks (PINNs) to reconstruct not only velocity but also hidden fields such as pressure from sparse experimental or simulated measurements, an advance that unlocks detailed fluid-structure interaction analyses for applications ranging from underwater vehicles to aerospace systems.

To date, most super‐resolution efforts have targeted individual macroscopic fields or treated quantities such as pressure and vorticity, separately. Our approach departs from this paradigm by operating on the lattice Boltzmann mesoscopic density distribution functions instead of the derived macroscopic variables. Because each lattice site carries multiple directional populations, these distributions embed richer flow dynamics and offer more degrees of freedom for the network to learn. Consequently, the resulting model achieves both superior accuracy and enhanced generalization ability across flow regimes. An additional benefit is that a single trained network suffices, i.e., once the high‐resolution distributions are reconstructed, one can derive velocity, pressure, vorticity and density simultaneously without retraining. This unified framework streamlines computation and preserves the inherent coupling among flow variables, a feature that is especially valuable for multiphysics simulations requiring concurrent access to diverse flow fields.

The remainder of this paper is organized as follows. In Sec.~\ref{sec:method}, we introduce the proposed super-resolution lattice Boltzmann method (SRLBM) and detail the deep neural network architectures used to upscale mesoscopic distribution functions across multiple scales. Section~\ref{sec:train} describes the high-fidelity dataset associated with a flow past a cylinder, outlines the data pre-processing and augmentation procedures, and presents the training and validation results. In Sec.~\ref{sec:superes}, we apply SRLBM to reconstruct macroscopic flow fields, including velocity, pressure, and vorticity, over various upscaling factors and assess its generalization ability to more complex two-cylinder configurations. Finally, Section~\ref{sec:conclude} summarizes our key findings and discusses future directions.

\section{\label{sec:method}Super-resolution lattice Boltzmann method (SRLBM)}

\begin{figure*}
  \centerline{\includegraphics{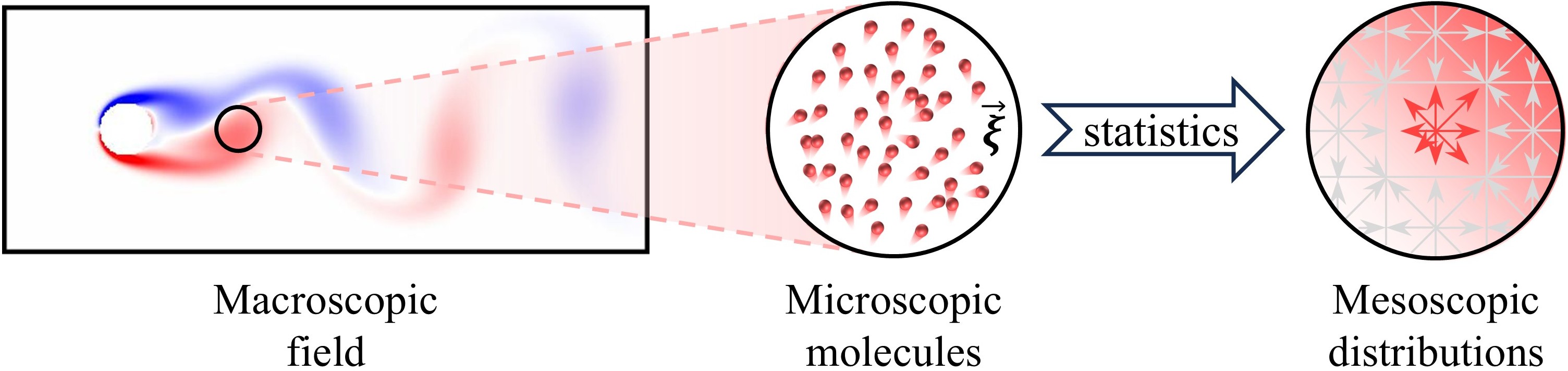}}
  \caption{Multi-scale description of fluid flows. The macroscopic flow field is essentially a collective behavior of microscopic molecules, which can also be represented by a set of mesoscopic density distribution functions through statistical averaging.}
  \label{fig:multiscale}
\end{figure*}

Incompressible flows, most often described by macroscopic fields of velocity and pressure, arise from the collective motion and density fluctuations of countless molecules, i.e., entities whose individual paths are impossible to track. Kinetic theory provides a bridge between these scales by defining mesoscopic distribution functions, which are statistical averages of molecular velocities that retain key information about momentum transfer and density variation. By operating at this intermediate mesoscopic level, one can capture intricate flow phenomena, such as eddy formation and wave propagation, without resorting to full molecular dynamics (see Fig.~\ref{fig:multiscale}). This multiscale perspective underpins the super-resolution lattice Boltzmann method proposed here.

\subsection{\label{subsec:lbm}Lattice Boltzmann method}

The lattice Boltzmann method describes fluid dynamics through a discrete set of mesoscopic distribution functions $f_i(\bm{x},t)$. Each $f_i$ represents the mass density of particles at position $\bm{x}$ and time $t$ traveling with one of the prescribed discrete velocities $\bm{c}_i$. Under the Bhatnagar-Gross-Krook (BGK) approximation \citep{bhatnagarModelCollisionProcesses1954}, collisions relax each $f_i$ toward its local equilibrium $f_i^{eq}$ over a single relaxation time $\tau$. The resulting lattice Boltzmann equation combines collision and streaming in a single update 
\begin{equation}
    f_i(\bm{x}+\bm{c}_i \delta_t, t+\delta_t) - f_i(\bm{x}, t) = -\frac{1}{\tau} \left[ f_i(\bm{x}, t) - f_i^{eq}(\bm{x}, t) \right],
    \label{eq:lbe}
\end{equation}
where $\delta_t$ is the time step. The equilibrium distribution function $f_i^{eq}$ depends on the local macroscopic fluid density $\rho$ and velocity $\bm{u}$ \citep{qianLatticeBGKModels1992}
\begin{equation}
    f_i^{eq}(\rho, \bm{u}) = w_i \rho \left[ 1 + \frac{\bm{c}_i \cdot \bm{u}}{c_s^2} + \frac{(\bm{c}_i \cdot \bm{u})^2}{2c_s^4} - \frac{\bm{u}^2}{2c_s^2} \right].
    \label{eq:feq}
\end{equation}

Following \citet{qianLatticeBGKModels1992}, the D2Q9 lattice structure is adopted in this study for two-dimensional fluid flow simulations. As a result, the speed of sound in lattice units is $1/\sqrt{3}$ and the weighting coefficients $w_i$ become $w_0 = 4/9$ associated with the discrete velocity at rest, $w_i = 1/9$ for $i = 1 \sim 4$ streaming in the orthogonal directions and $w_i = 1/36$ for $i = 5 \sim 8$ streaming in the diagonal directions. Macroscopic flow quantities, including density and velocity, can be recovered by taking the first and second velocity moments of the density distribution functions according to the conservation of mass and momentum
\begin{align}
    \rho(\bm{x}, t) &= \sum_{i=0}^8 f_i(\bm{x}, t) \, , \\
    \bm{u}(\bm{x}, t) &= \frac{1}{\rho} \sum_{i=0}^8 f_i(\bm{x}, t) \bm{c}_i \, .
\end{align}

Through a multiscale Chapman-Enskog expansion, the Navier-Stokes equations can be derived from the lattice Boltzmann equation, i.e., Eq.~\eqref{eq:lbe}, resulting in a relationship between the fluid kinematic viscosity $\nu$ and the relaxation time $\tau$ \citep{heLatticeBoltzmannModel1997}
\begin{equation}
    \nu = c_s^2 \left( \tau - \frac{1}{2} \right) \frac{\delta_x^2}{\delta_t} \, ,
\end{equation}
where $\delta_x$ is the size of a lattice cell. A weak fluid compressibility is allowed in the single relaxation time LBM and the macroscopic fluid pressure can be obtained from the equation of state $p = c_s^2 \rho$.

The particulate nature of LBM is considered to be advantageous for simulating fluid-structure interactions. In this study, the partially saturated cell (PSC) method originally proposed by \citet{nobleLatticeBoltzmannMethodPartially1998} has been adopted to treat solid obstacles inside a flow. The PSC method generalizes the lattice Boltzmann equation by introducing a new collision operator that is capable of describing the fluid and solid behaviors in a unified manner
\begin{equation}
    f_i(\bm{x}+\bm{c}_i \delta_t, t+\delta_t) - f_i(\bm{x}, t) = \left[ 1 - B(\phi, \tau) \right] \Omega_i^f + B(\phi, \tau) \Omega_i^s \, ,
    \label{eq:psc}
\end{equation}
where $B(\phi, \tau)$ is a factor that weighs the relative importance of fluid and solid dynamics of a lattice cell, which depends on the solid volume fraction $\phi$ and the relaxation time $\tau$
\begin{equation}
    B(\phi, \tau) = \frac{\phi(\tau - 1/2)}{(1 - \phi) + (\tau - 1/2)} \, .
\end{equation}

The weighting factor $B$ increases from 0 to 1 as the solid volume fraction $\phi$ increases from 0 to 1. The fluid collision operator $\Omega_i^f$ takes the form of BGK, i.e., the right hand side of Eq.~\eqref{eq:lbe}. The solid collision operator $\Omega_i^s$ is defined with the bounce back of the non-equilibrium density distribution functions to ensure no-slip between fluid and solid
\begin{equation}
    \Omega_i^s = f_{-i}(\bm{x}, t) - f_{-i}^{eq}(\rho, \bm{u}) + f_i^{eq}(\rho, \bm{u}_s) - f_i(\bm{x}, t) \, ,
\end{equation}
where $\bm{u}_s$ denotes the solid velocity at the position of the lattice cell $\bm{x}$, which is set to zero since in this study no moving solid is considered. The subscript $-i$ represents the opposite direction of $i$.

The PSC method preserves the locality of the nonlinear collision operator while retaining the linearity of the nonlocal streaming process. By allowing the weighting factor $B$ to continuously vary between 0 and 1, partially saturated cells produce a more accurate fluid-solid interface than the staircase boundary that results from a traditional bounce-back scheme, which usually places the interface midway between neighboring lattice nodes.

\subsection{\label{subsec:sr}Multiscale super-resolution neural networks}

\begin{figure*}
  \centerline{\includegraphics{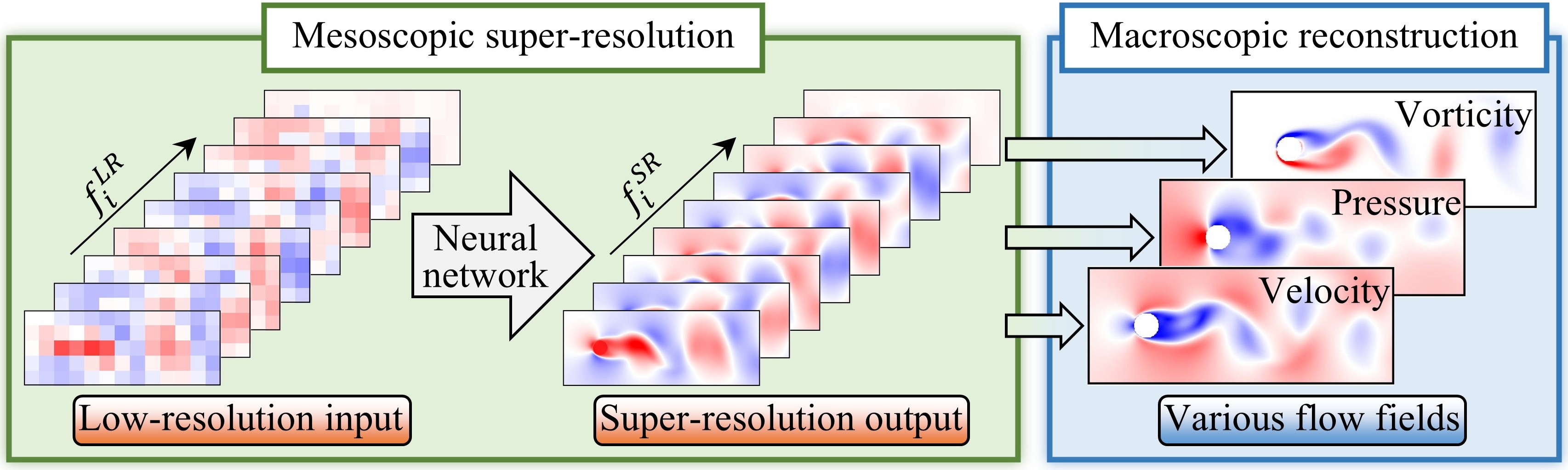}}
  \caption{Multiscale super-resolution reconstruction of macroscopic fluid flows based on the mesoscopic density distribution functions from lattice Boltzmann simulations.}
  \label{fig:sr}
\end{figure*}

We use supervised learning to train our neural networks for flow super‐resolution, which requires sufficient paired low‐resolution inputs and high‐resolution outputs. In the lattice Boltzmann method, each fluid cell recovers its macroscopic properties from mesoscopic density distribution functions. Consequently, rather than directly super‐resolving macroscopic fields such as velocity or vorticity, as in previous studies \citep{fukamiSuperresolutionReconstructionTurbulent2019, liuDeepLearningMethods2020}, we enhance the resolution of density distribution functions (see Fig.~\ref{fig:sr}) by
\begin{equation}
    f_i^{SR} = \mathcal{F}(f_i^{LR}, \theta) \, ,
    \label{eq:sr}
\end{equation}
where $f_i^{LR}$ and $f_i^{SR}$ represent the low-resolution input and the high-resolution output, respectively. All learnable parameters of a neural network are denoted by $\theta$. The neural network works as a nonlinear mapping function $\mathcal{F}$. Then, the training process becomes essentially an optimization task such that
\begin{equation}
    \theta = \underset{\theta}{\arg\min} \, \mathcal{L}(\mathcal{F}(f_i^{LR}, \theta), f_i^{HR}) \, ,
    \label{eq:optim}
\end{equation}
where $f_i^{HR}$ represents the high-resolution reference data and $\mathcal{L}$ is the loss function. Once the training process completes and the loss converges to a sufficiently small value, the optimized $\theta$ can be substituted into Eq.~\eqref{eq:sr} for fast super-resolution reconstructions of density distribution functions. After that, the super-resolution versions of the macroscopic density ($\rho^{SR}$) and velocity ($\bm{u}^{SR}$) fields can be obtained
\begin{align}
    \rho^{SR} &= \sum_{i=0}^8 f_i^{SR} \, , \\
    \bm{u}^{SR} &= \frac{1}{\rho} \sum_{i=0}^8 f_i^{SR} \bm{c}_i \, .
\end{align}

And $\rho^{SR}$ can be further applied to reconstruct the super-resolution pressure ($p^{SR}$) field according to the equation of state. The super-resolution vorticity field can be calculated from the corresponding velocity field by using a second-order finite difference scheme.

We denote this multiscale super‐resolution flow reconstruction framework as the super‐resolution lattice Boltzmann method (SRLBM). In SRLBM, the state of each lattice cell is represented by a vector of mesoscopic density distribution functions, an arrangement directly analogous to a multi-channel image pixel, which combines multiple color channels (e.g. red, green, blue) into a single visual value \citep{chenCompressedLatticeBoltzmann2021}. This formal equivalence allows many existing image super‐resolution architectures to be adapted for SRLBM with only minimal changes to their input and output channel dimensions. In this work, we implement and compare two such architectures, a standard convolutional neural network (CNN) and a deeper residual dense network (RDN), to evaluate their relative performance in reconstructing high‐fidelity flow fields across scales. Furthermore, in the present framework, boundary conditions are implicitly embedded in data rather than by modifying the inner structure of neural networks. Therefore, SRLBM can be trained on data with arbitrary boundary setups and it generalizes to any combination of boundary conditions, e.g. Dirichlet, Neumann, mixed and moving walls, without architectural changes. In contrast, super-resolution methods that rely on frequency domain decompositions inherently assume periodicity in each coordinate direction to satisfy basis function orthogonality, and therefore cannot be directly applied to non-periodic or mixed-type boundary conditions without additional problem-specific treatment \citep{xiongHighfrequencyFlowField2025}.

\subsubsection{\label{subsubsec:cnn}Convolutional neural network (CNN)}

\begin{figure*}
  \centerline{\includegraphics{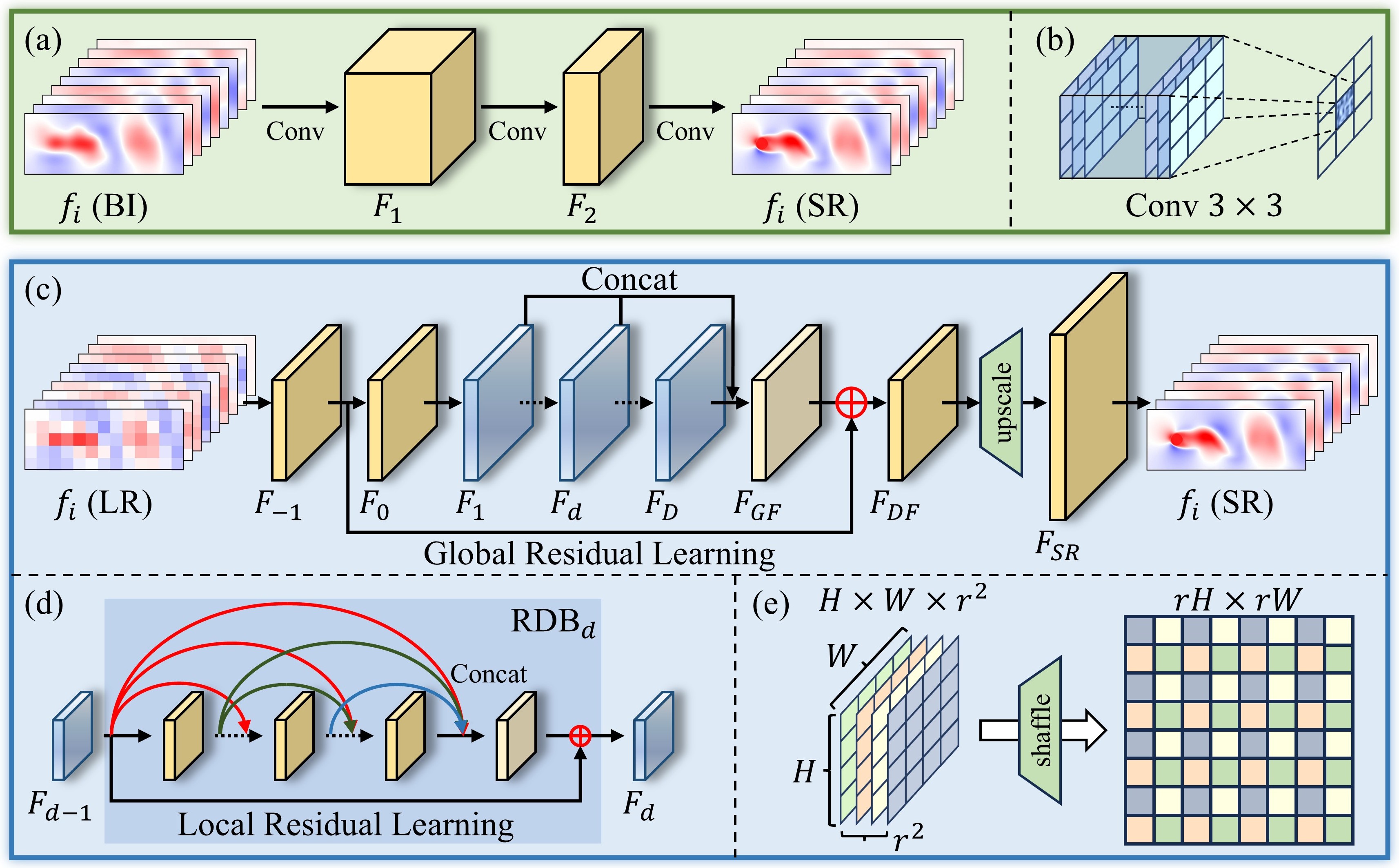}}
  \caption{(a) Architecture for super-resolution reconstruction of density distribution functions based on the convolutional neural network. (b) Schematic of convolution (Conv) operation with a kernel size of 3. (c) Architecture for super-resolution reconstruction of density distribution functions based on the residual dense network. (d) Neural network architecture of a residual dense block. (e) Schematic of the PixelShuffle operation that transforms a tensor of the shape $H \times W \times r^2$ into another tensor of the shape $rH \times rW$.}
  \label{fig:nn}
\end{figure*}

Figure~\ref{fig:nn}(a) depicts the CNN architecture we employ, which consists of two sequential convolutional layers. To generate the low-resolution (LR) input, we first apply bicubic interpolation (BI) to the high-resolution (HR) density distribution functions, intentionally degrading fine-scale details while preserving the original spatial dimensions. Zero padding is used at each convolutional stage to maintain this fixed size throughout the network.

Under the D2Q9 lattice model (see Sec.~\ref{subsec:lbm}), the state of each lattice cell comprises nine distribution functions, one for each discrete velocity direction, so the native data naturally form a 9-channel tensor. To provide the network with crucial geometric information, we add a tenth channel encoding the local solid volume fraction, thus distinguishing fluid from solid regions. Thus, the CNN accepts a 10-channel LR tensor and produces 9 super-resolved distribution channels, in contrast to conventional image super-resolution networks that typically handle one (grayscale) or three (RGB) channels. Note that the input channel for the solid volume fraction is not shown in Fig.~\ref{fig:nn} for simplicity.

Following \citet{dongImageSuperResolutionUsing2016}, our CNN begins with a first convolutional layer (Conv1) that applies a set of learnable filters to the low-resolution input, extracting a feature tensor $F_1$ that encodes local flow patterns. A second convolutional layer (Conv2) then nonlinearly maps $F_1$ into high-resolution patch representations $F_2$. Figure~\ref{fig:nn}(b) illustrates a single $3 \times 3$ convolution operation, and the rectified linear unit (ReLU) activation follows each convolution to introduce nonlinearity and promote sparse, stable feature learning \citep{glorotDeepSparseRectifier2011}. Table~\ref{tab:cnn} summarizes the detailed CNN configuration, including the number of input and output channels, kernel sizes, and padding values for each layer.

\begin{table*}
\caption{\label{tab:cnn}Detailed settings for the convolutional neural network. The height and width of the patches randomly sampled from the domain for super-resolution are denoted by $H$ and $W$, respectively.}
\begin{ruledtabular}
  \begin{tabular}{lccccc}
    Layer  & Input channels & Output channels & Kernel size & Padding & Output size             \\
    \hline
    Input  & $\sim$         & $\sim$          & $\sim$      & $\sim$  & $H \times W \times 10$  \\
    Conv1  & 10             & 64              & 9           & 4 & $H \times W \times 64$        \\
    Conv2  & 64             & 32              & 5           & 2 & $H \times W \times 32$        \\
    Output & 32             & 9               & 5           & 2 & $H \times W \times 9$         \\
  \end{tabular}
\end{ruledtabular}
\end{table*}

Despite its success in single-image super-resolution, the basic CNN architecture exhibits two critical shortcomings when applied to flow fields. First, it relies on pre-upsampling the low-resolution input, often via bicubic interpolation, to the target grid, a step that inherently smooths fine-scale details and raises computational cost roughly quadratically \citep{dongAcceleratingSuperResolutionConvolutional2016}. Second, its strictly sequential stacking of convolutional layers means that only the deepest feature maps inform the final reconstruction, while the rich, multi-scale representations captured in intermediate layers are discarded. By ignoring this hierarchical information, which is vital for flows that span from small eddies to large coherent structures, the standard CNN may struggle to recover the full spectrum of scales, especially when the flow becomes complex (see Sec.~\ref{sec:superes}).

\subsubsection{\label{subsubsec:rdn}Residual dense network (RDN)}

Figure~\ref{fig:nn}(c) presents the RDN architecture proposed by \citet{zhangResidualDenseNetwork2018}, which effectively addresses the shortcomings of a plain CNN. The network comprises four components: the shallow feature extraction network (SFENet), residual dense blocks (RDBs), dense feature fusion (DFF) and the up‐sampling network (UPNet). In SFENet, two convolutional layers extract shallow features $F_{-1}$ and $F_{0}$. The first feature map $F_{-1}$ serves in global residual learning, while the second map $F_{0}$ is fed into the RDBs.

Let $D$ be the total number of RDBs. We denote the input to the $d$th block by $F_{d-1}$ and its output by $F_{d}$. Each RDB shares the same architecture, which is a sequence of densely connected convolutional layers, see Fig.~\ref{fig:nn}(d). Within a block, every convolutional layer takes as input the concatenation of $F_{d-1}$ and all feature maps produced by the preceding layers in that block. Each output then passes through a ReLU activation. After these layers, local feature fusion (LFF) applies a $1 \times 1$ convolution to merge the original input $F_{d-1}$ with the intermediate feature maps. Finally, a local residual learning shortcut adds $F_{d-1}$ element-wise to the fused features, enhancing the representational capacity of the block, as demonstrated in deep residual learning \citep{heDeepResidualLearning2016}. The output of the final block is denoted by $F_{D}$.

After the RDBs extract local dense features, dense feature fusion (DFF) merges these multi-scale representations. First, the outputs of all RDBs are concatenated and then passed through a $1 \times 1$ convolution to compress channels, followed by a $3 \times 3$ convolution to blend spatial information. This sequence, referred to as global feature fusion (GFF), produces the global feature map $F_{GF}$. Finally, global residual learning (GRL) combines $F_{GF}$ with the initial shallow feature map $F_{-1}$ via element-wise addition, yielding the dense feature map $F_{DF}$.

Finally, we employ a sub‐pixel convolutional layer, often called PixelShuffle \citep{shiRealTimeSingleImage2016}, as the up‐sampling network (UPNet). This layer rearranges the dense feature map $F_{DF}$ of size $H\times W\times n\times r^2$ into a high‐resolution feature tensor $F_{SR}$ of size $(rH)\times(rW)\times n$, where $r$ is the upscaling factor and $n$ is the number of feature channels. Figure~\ref{fig:nn}(e) illustrates how PixelShuffle reshapes and permutes these dimensions. A final convolutional layer then reconstructs the super‐resolved density distribution functions. Table~\ref{tab:rdn} details the layer configurations of RDN. As with the CNN, the RDN takes 9 density distribution channels ($f_0$ through $f_8$) plus a solid volume fraction channel, i.e., 10 inputs in total, and produces 9 channels corresponding to the super‐resolved distributions.

\begin{table}
\caption{\label{tab:rdn}Detailed settings for the residual dense network.}
\begin{ruledtabular}
    \begin{tabular}{lc}
    Parameters                                          & Value         \\
    \hline
    Input channels                                      & 10            \\
    Number of RDBs                                      & 16            \\
    Number of convolutional layers in RDB               & 8             \\
    Number of features                                  & 64            \\
    Kernel size of convolutional layers for LFF and GFF & $1\times1$    \\
    Kernel size of other convolutional layers           & $3\times3$    \\
    Output channels                                     & 9             \\
  \end{tabular}
\end{ruledtabular}
\end{table}

Unlike the plain CNN, which begins by up-sampling its input, the RDN operates directly on low-resolution data, eliminating costly interpolation and streamlining the entire super-resolution workflow. Its densely connected residual blocks produce a rich hierarchy of local features, from the smallest vortices to the largest flow patterns, and then fuse them seamlessly. By fully exploiting this multiscale feature space, RDN aligns naturally with the scale-spanning character of complex flows, yielding more faithful and detailed high-fidelity flow reconstructions.

\subsubsection{\label{subsubsec:optim}Network optimization}

We train both CNN and RDN with the adaptive moment estimation (Adam) optimizer \citep{fukamiSuperresolutionReconstructionTurbulent2019}, adjusting all network parameters to minimize a scalar loss that measures the discrepancy between super‐resolved outputs and high‐resolution ground truth. Although CNNs traditionally use mean square error (MSE) loss \citep{dongImageSuperResolutionUsing2016} and RDNs often employ an $L_1$ norm \citep{zhangResidualDenseNetwork2018}, our experiments showed a negligible difference in flow reconstruction accuracy between the two for test cases presented in this study. For consistency and stable convergence, we therefore adopt the MSE loss throughout. Formally, if $\tilde{f}_i^{SR}$ and $\tilde{f}_i^{HR}$ denote the normalized super‐resolved and reference distribution functions at each lattice direction $i$, the loss becomes
\begin{equation}
    \mathcal{L}(\theta) = \frac{1}{N} \sum_{k=1}^N \left( \sum_{i=0}^8 \lVert \tilde{f}_i^{SR}\left( \theta \right) - \tilde{f}_i^{HR}\rVert^2 \right)_k \, ,
    \label{eq:optim}
\end{equation}
where $N$ is the number of training samples and $k$ is the counting index. To minimize the loss function, an initial learning rate of $1\times10^{-4}$ is adopted. During training, the learning rate will be reduced by a factor of 10 if there is no improvement after 10 consecutive training epochs.

\section{\label{sec:train}Data acquisition and model training}

\subsection{\label{subsec:data}Data generation}

\begin{figure*}
  \centerline{\includegraphics{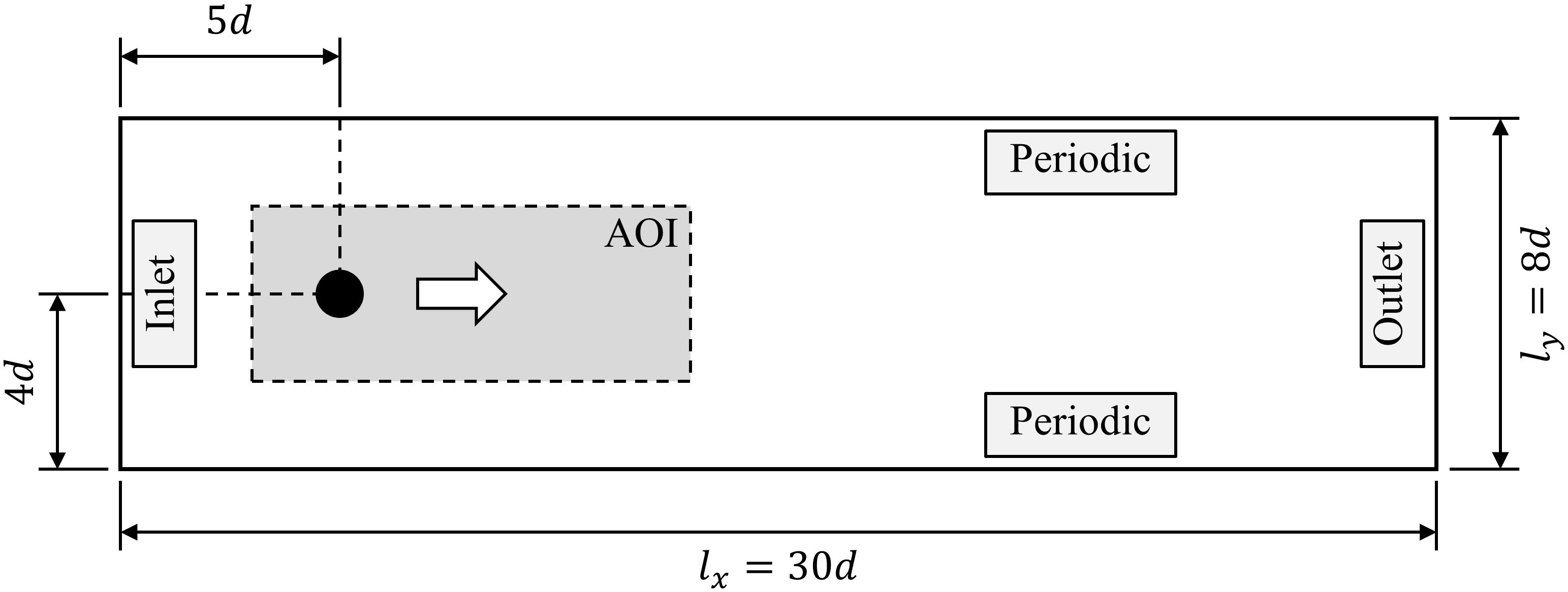}}
  \caption{Sketch of the numerical setup for the lattice Boltzmann simulation of a flow past a fixed cylinder (black circle). The gray area around the cylinder shows the area of interest (AOI), where the flow information is extracted for the training and testing of the machine learning models.}
  \label{fig:setup}
\end{figure*}

To generate the training dataset, we perform high-fidelity lattice Boltzmann simulations of a two-dimensional flow past a circular cylinder, see Fig.~\ref{fig:setup}. We denote the cylinder diameter by $d$, and set the computational domain to a streamwise length $l_x = 30d$ and $l_y = 8d$ in the cross-stream direction. A uniform inlet velocity $U$ is imposed on the left boundary, the right boundary enforces a zero-gradient outlet condition, and periodic boundaries are applied on the top and bottom to minimize confinement effects. The cylinder sits on the horizontal centerline, $5d$ downstream of the inlet, allowing a fully developed wake to form. We choose a spatial resolution of $d / \delta_x = 32$ and a relaxation time of $\tau = 0.53$ to ensure that the wake dynamics is well resolved. Extensive validation against reference drag coefficients, pressure distributions, recirculation lengths, and flow-separation angles at Reynolds numbers $\mathrm{Re}=Ud/\nu = 20$ and 40 confirms excellent agreement with the results of the literature \citep{LuoBoltzmann2024}. Note that the steady-state results at $\mathrm{Re}=20$ and 40 are only used for the verification of numerical calculations, which are not presented here.

\begin{figure*}
  \centerline{\includegraphics{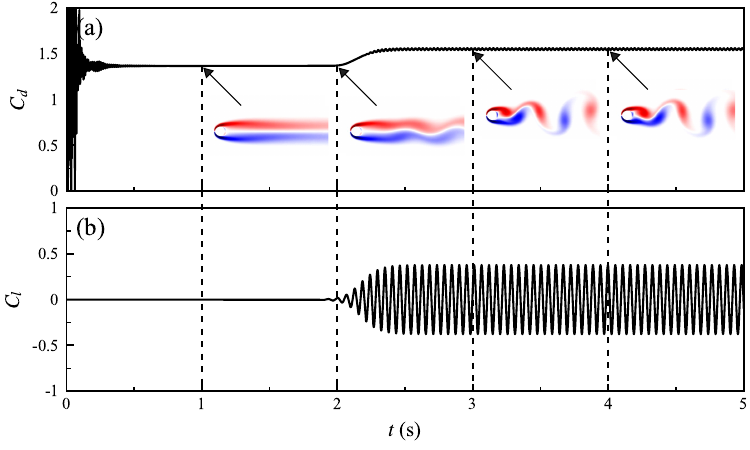}}
  \caption{Results of the lattice Boltzmann simulation of a flow past a fixed cylinder when $\mathrm{Re} = 100$. (a) Temporal evolution of the drag coefficient. The insets show the vorticity fields around the cylinder at $t = 1$ s, 2 s, 3 s and 4 s. (b) Temporal evolution of the lift coefficient.}
  \label{fig:coeffs}
\end{figure*}

We train both CNN and RDN on data from a fully resolved LBM simulation at $\mathrm{Re} = 100$, where vortex shedding produces rich multiscale structures. Although the complete grid spans $960 \times 256$ lattice cells, we extract only the region around the cylinder, resulting in an area of interest (AOI) measuring $13d \times 5d$ ($416 \times 160$ lattice cells, see Fig.~\ref{fig:setup}), to focus learning on the most dynamically active flow. At $\mathrm{Re} = 100$, the flow transitions to an unsteady regime with periodic vortex shedding once fully developed. We monitor the drag coefficient $C_d = F_x/(0.5\rho U^2 d \delta_x)$ and the lift coefficient $C_l = F_y/(0.5\rho U^2 d \delta_x)$, where $F_x$ and $F_y$ are the hydrodynamic forces on the cylinder in the streamwise and spanwise directions, respectively. Until $t=2$ s, the wake remains symmetric about the horizontal centerline, and $C_l$ is near zero (see the inset vorticity at $t=1$ s in Fig.~\ref{fig:coeffs}). Beyond $t\approx2$ s, the symmetry breaks, $C_l$ oscillates between positive and negative peaks, and $C_d$ rises to its established mean, consistent with previous studies \citep{rajaniNumericalSimulationLaminar2009}. We therefore select unsteady snapshots from $t=2$ s to $t=5$ s as our training set, ensuring that the networks learn to reconstruct both small-scale local vortices and large‐scale wake patterns.

To enrich our training set without extra simulations, we have applied systematic data augmentation. From each unsteady snapshot, we extract 20 random $64 \times 64$ cell patches within the AOI. Each patch then undergoes random horizontal and vertical flips plus $90^{\circ}$ rotations, generating geometric variants that preserve the core flow physics. These transformations expose the networks to shifted, reflected, and reoriented vortex patterns, imposing invariance to translations and symmetries in the flow. As a result, model robustness improves and overfitting diminishes, since the networks learn to recognize essential flow features under diverse geometric configurations \citep{shortenSurveyImageData2019}.

Low-resolution inputs are generated by bicubic downsampling of the high-resolution distribution functions at upscaling factors $r=2$, 4, and 8. This degradation mimics the smooth loss of fine-scale detail while preserving overall flow structure, following common image SR practices. For each scale, we assemble paired LR--HR patches and then split them randomly into training (80\%) and testing (20\%) sets. This partitioning ensures that the networks learn robust multiscale mappings while providing an unbiased evaluation of super-resolution performance.

\subsection{\label{subsec:loss}Training results}

\begin{figure}
  \centerline{\includegraphics{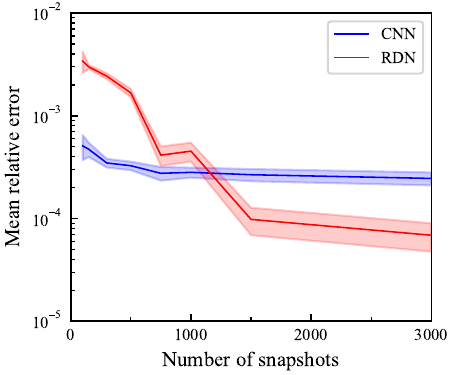}}
  \caption{Plot of the mean relative errors of the CNN and RDN models against the number of snapshots at the super-resolution scale of 8. The error is calculated based on the testing data. The shaded area indicates the range of variations set by three times of the standard deviations.}
  \label{fig:train_snapshots}
\end{figure}

To ensure sufficient training coverage, we extract varying numbers of snapshots evenly from $t=2$ s to 5 s within the AOI. Figure~\ref{fig:train_snapshots} plots the mean relative error $\left| (f_i^{SR} - f_i^{HR})/f_i^{HR} \right|$ of the test set against the number of snapshots for both CNN and RDN. We favor this normalized metric because each distribution $f_i$ scales with its lattice weight $w_i$, see Eq.~\eqref{eq:feq}, making absolute errors in the dominant rest population $f_0$ disproportionately large. The shaded bands mark three times the standard deviation in the testing samples. In the low‐data regime (fewer than 1000 snapshots), the simpler CNN achieves lower error than the deeper RDN, since its compact architecture requires less data to stabilize. As we increase the dataset, the error of CNN quickly plateaus, while the error of RDN continues to drop sharply until about 1500 snapshots, after which the gains diminish. At 3000 snapshots, the RDN achieves a mean relative error of $6.88\times10^{-5}$, significantly below the error of CNN, i.e., $2.46\times10^{-4}$. Training neural networks with small datasets can be extremely powerful, e.g., by extracting rich flow information from a single turbulent snapshot \citep{fukamiSinglesnapshotMachineLearning2024} or embedding physical priors directly into the learning process \citep{pageSuperresolutionTurbulenceDynamics2025}, one can dramatically reduce the data requirements for accurate reconstruction. Furthermore, both CNN and RDN architectures can be flexibly scaled, by tuning layer depth, feature channel widths, or the number of residual blocks, to find the optimal balance between reconstruction accuracy and computational efficiency under different data budgets. In this study, however, our objective is to benchmark the peak performance of both models, so we train each on a sufficiently large dataset of 3000 snapshots to ensure we evaluate their capabilities at the performance ceiling.

\begin{figure*}
  \centerline{\includegraphics{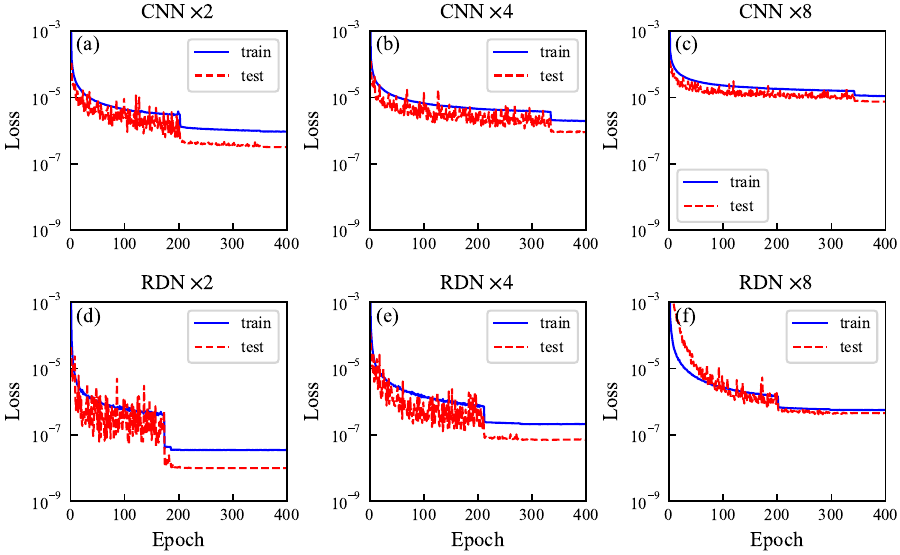}}
  \caption{Learning curves for different models and upscaling factors using 3000 snapshots evenly extracted from $t=2$ s to 5 s. (a-c) CNN model with upscaling factors of 2, 4 and 8; (d-f) RDN model with upscaling factors of 2, 4 and 8.}
  \label{fig:train_loss}
\end{figure*}

Figure~\ref{fig:train_loss} plots the MSE loss trajectories, optimized through Eq.~\eqref{eq:optim}, for both CNN and RDN at upscaling factors $r=2$, 4, and 8. All curves stabilize by the 400th epoch, indicating convergence of the training process. The final test loss remains consistently smaller than the training loss, which is caused by the heavy data augmentations during training. As the super-resolution scale increases, the asymptotic loss rises, reflecting the growing challenge of reconstructing finer flow details from coarser inputs. Crucially, at every scale the training and validation losses of RDN remain over an order of magnitude lower than those of the CNN, underscoring the superior capacity of RDN to capture the multiscale features inherent in fluid flows.

\begin{figure*}
  \centerline{\includegraphics{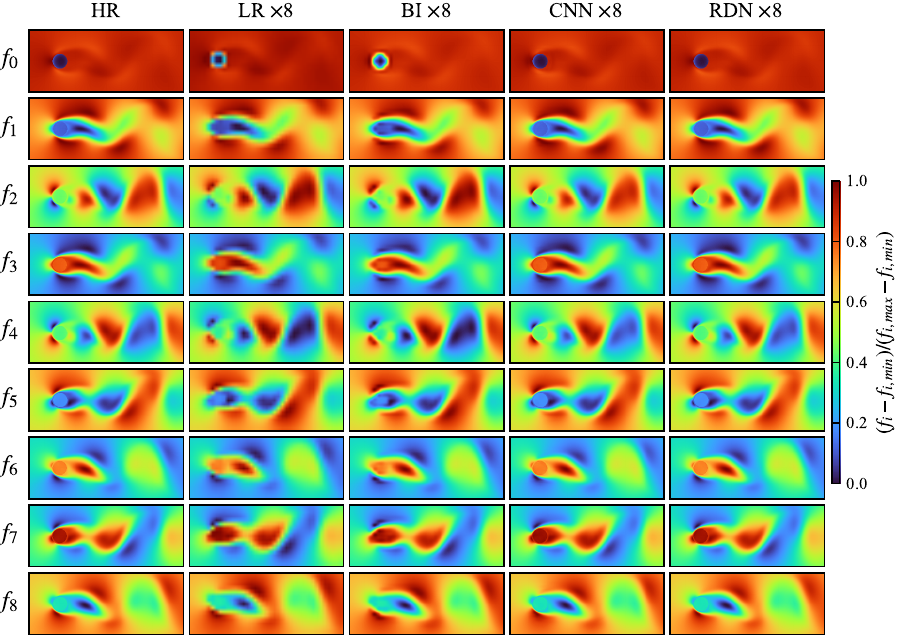}}
  \caption{Comparison of the density distribution functions around the cylinder at a randomly selected time instant in the testing set. The condition with the upscaling factor of 8 is considered. First column: high-resolution (HR) reference data, second column: low-resolution (LR) input data, third column: output data from bicubic interpolation (BI), fourth column: super-resolution output data from CNN, fifth column: super-resolution output data from RDN.}
  \label{fig:train_dfs}
\end{figure*}

Figure~\ref{fig:train_dfs} compares the nine distribution functions $f_0$ to $f_8$ around the cylinder at a randomly selected time instant in the test set for an upscaling factor $r=8$. For each $f_i$, we display the high-resolution ground truth, the low-resolution input, its bicubic interpolation, and the super-resolved output from CNN and RDN. All fields are Min--Max normalized to the range $[0,1]$ for consistent visualization. Under such extreme downsampling, i.e., only four grid points span the cylinder diameter, the LR fields become heavily pixelated, and BI merely smooths the edges while losing fine details and blurring steep transitions at the fluid-solid interface (third column of Fig.~\ref{fig:train_dfs}).

In contrast, both CNN and RDN reconstructions (fourth and fifth columns of Fig.~\ref{fig:train_dfs}) restore sharp cylinder boundaries and recover coherent wake patterns that closely match the HR reference. This dramatic improvement arises because the networks learn nonlinear correlations that infer subgrid-scale fluctuations from their surroundings. The dense residual connections of RDN, in particular, enhance subtle gradient transitions and sharpen small‐scale features. These results underscore that the learned super-resolution transcends simple pixel interpolation. Overall, the machine learning models are able to reconstruct multiscale physics encoded in density distribution functions effectively.

\section{\label{sec:superes}Multiscale super-resolution flow reconstruction}

Extending super-resolution to the kinetic realm of density distribution functions $f_i$, rather than operating on derived macroscopic variables, we obtain two transformative benefits. First, the lattice distributions span a finer granularity, with each cell comprising nine directional populations, so the networks inherit richer flow physics from vortices to subtle shear layers. Embedding this mesoscopic detail into the training lets the models infer sharper, more physically consistent reconstructions. Second, once we predict the full set of $f_i$, we can derive multiple macroscopic quantities, such as velocity, pressure, vorticity, without retraining separate networks for each variable. In what follows, we demonstrate this multiscale framework on flow past a single cylinder, highlighting its reconstruction fidelity, and on two‐cylinder configurations, probing the ability of the models to generalize across more complex wake interactions.

\subsection{\label{subsec:onecylinder}Flow past one cylinder}

\begin{figure*}
  \centerline{\includegraphics{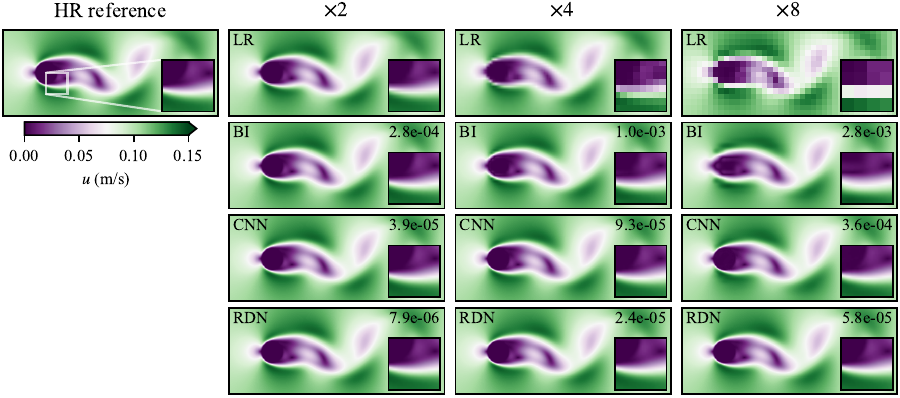}}
  \caption{Velocity field of the flow past a cylinder. First column: high-resolution (HR) velocity field. Second column: low-resolution (LR), bicubic interpolated (BI) and super-resolution (CNN and RDN) velocity fields when the upscaling factor is 2. Third column: LR, BI, CNN and RDN velocity fields when the upscaling factor is 4. Fourth column: LR, BI, CNN and RDN velocity fields when the upscaling factor is 8. The number at the top-right corner indicates the root-mean-square error if applicable. A zoom-in view of a region close to the cylinder is shown as an inset at the right-bottom corner. All subplots share the same colorbar.}
  \label{fig:srlbm_vel}
\end{figure*}

Figure~\ref{fig:srlbm_vel} presents the high-resolution, low-resolution, and super-resolved velocity fields for upscaling factors $r=2$, 4, and 8. In each panel, the root mean square error (RMSE) between the reconstructed and the HR fields is annotated at the top-right corner if applicable, while a zoomed inset focuses on the steep velocity gradients adjacent to the cylinder. As $r$ increases, all methods experience growing errors, reflecting the progressive loss of fine-scale structures in coarser inputs. However, both the CNN and the RDN models outperform bicubic interpolation at every scale. RDN excels in general, achieving the lowest RMSE, likely due to its deeper architecture and dense residual connections that capture multiscale features more effectively, see Sec.~\ref{sec:method}. Crucially, bicubic interpolation blurs the sharp velocity jump at the fluid-solid interface, especially at $r=8$, while CNN and RDN preserve sharp boundary layers, with RDN delivering the most faithful restoration of wake dynamics.

\begin{figure*}
  \centerline{\includegraphics{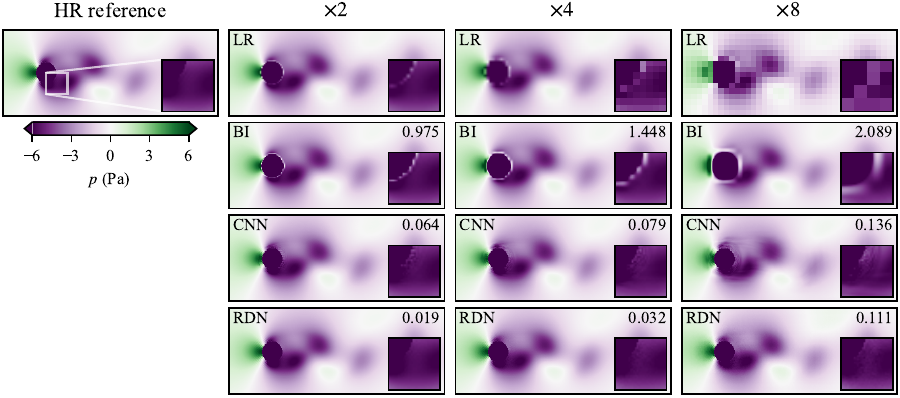}}
  \caption{Pressure field of the flow past a cylinder. The same arrangements of subplots as Fig.~\ref{fig:srlbm_vel} are adopted.}
  \label{fig:srlbm_p}
\end{figure*}

Pressure, derived from the same set of density distribution functions as velocity, proves even more sensitive to downsampling artifacts. In Fig.~\ref{fig:srlbm_p}, we present the high-resolution, low-resolution, and reconstructed pressure fields across upscaling factors $r=2$, 4, and 8. As $r$ increases, bicubic interpolation incurs the largest errors, while RDN consistently delivers the most accurate reconstructions. At $r=2$, interpolation produces an unphysical white halo, implying zero pressure around the cylinder, whereas the true field features a positive stagnation peak upstream and negative suction in the wake. At $r=8$, CNN reconstruction exhibits spurious pressure oscillations absent in both reference data and RDN output. Although the RMSE of RDN (0.111) is slightly smaller than the RMSE of CNN (0.136) at $r=8$, its pressure contours, especially near the cylinder surface, align far more closely with the ground truth. This contrast underscores that global metrics such as RMSE can obscure critical local discrepancies, which are essential to accurately capture pressure-driven forces and boundary layer dynamics.

\begin{figure*}
  \centerline{\includegraphics{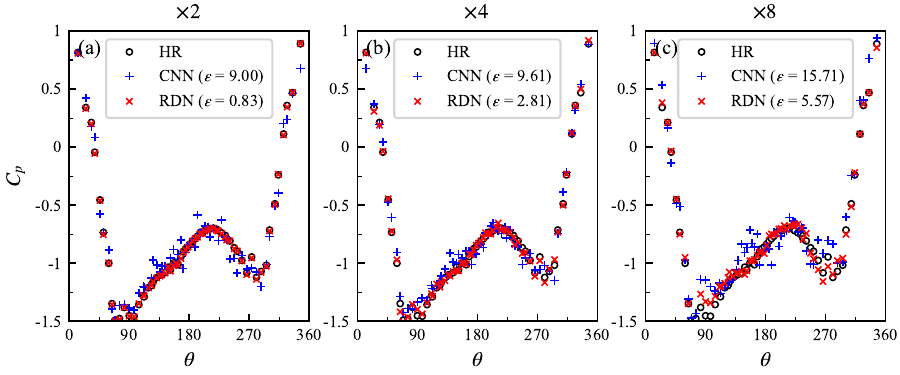}}
  \caption{Comparison of the high-resolution (HR) and super-resolution (CNN and RDN) pressure distributions around the cylinder wall at a randomly selected time instant in the testing dataset when $\mathrm{Re}=100$. (a) Upscaling factor of 2; (b) Upscaling factor of 4; (c) Upscaling factor of 8. The pressure coefficient $C_p$ equals to the pressure $p$ normalized by $0.5\rho U^2$.}
  \label{fig:srlbm_p360}
\end{figure*}

To further quantify reconstruction accuracy, we compute the pressure coefficient $C_p = p/(0.5\rho U^2)$ around the cylinder for both the high‐resolution reference and the super‐resolved fields. Figure~\ref{fig:srlbm_p360} plots $C_p$ as a function of the polar angle $\alpha$ (with $\alpha=0^\circ$ at the stagnation point and $\alpha=180^\circ$ on the lee side) for the upscaling factors $r=2$, 4, and 8. Bicubic interpolation is omitted here because of its excessively large deviations.

At $r=2$, RDN reproduces the entire pressure profile along the fluid-solid interface with remarkable fidelity. Its total absolute error $\varepsilon=0.83$ is an order of magnitude smaller than that of CNN ($\varepsilon=9.00$). In contrast, the suction peaks of CNN behind the cylinder exhibit spurious oscillations, revealing its limited ability to infer strong negative pressures. When $r=4$, the error of RDN grows noticeably at angles around $\alpha=90^\circ$ and $270^\circ$, where the true pressure gradient changes rapidly. Meanwhile, the overall error of CNN remains nearly unchanged, reflecting its systematic smoothing of sharp features. At the extreme scale $r=8$, both networks struggle as the oscillating wake induces complex pressure fluctuations across $\alpha\in(90^{\circ}, 270^{\circ})$, yet RDN still outperforms CNN in capturing the dominant trend. These results underscore the capacity of RDN to reconstruct critical pressure distributions even under severe downsampling, demonstrating the potential of SRLBM for efficient high-fidelity fluid-structure interaction simulations.

\begin{figure*}
  \centerline{\includegraphics{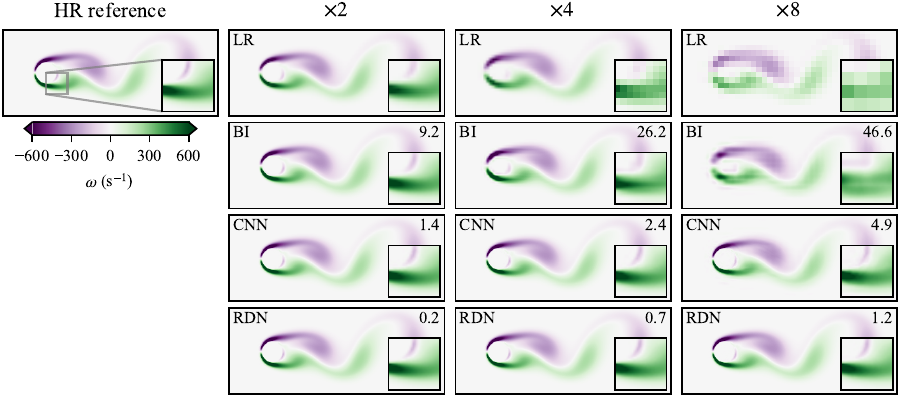}}
  \caption{Vorticity field of the flow past a cylinder. The same arrangements of subplots as Fig.~\ref{fig:srlbm_vel} are adopted.}
  \label{fig:srlbm_vor}
\end{figure*}

Another crucial macroscopic flow quantity is the vorticity, which is calculated based on the derivatives of the velocities, i.e., $\omega = \partial u_y / \partial x - \partial u_x / \partial y$, where $u_x$ and $u_y$ are the flow velocities in the $x$ and $y$ directions, respectively. According to \citet{fukamiSuperresolutionReconstructionTurbulent2019}, the vorticity field is amplified by the wavenumber compared to the velocity field, which stresses the presence of small scale flow structures. The vorticity fields reconstructed by various methods are compared in Fig.~\ref{fig:srlbm_vor}. Again, the reconstruction error increases as the upscaling factor increases, and RDN is the most accurate method compared to CNN and bicubic interpolation. Take the results with upscaling factor of 8 as an example, it is rather encouraging that the machine learning models are capable of restoring the major flow structures close to the cylinder, while the low resolution input appears to lose most of the flow information, see the last column of Fig.~\ref{fig:srlbm_vor} and the zoom-in views. Note that the frame is selected from the test set, which is not used for training the machine learning models. In contrast, the interpolation method cannot recover extreme negative and positive vorticity values around the cylinder, further demonstrating the effectiveness of the machine learning models.

\subsection{\label{subsec:twocylinders}Flow past two cylinders}

Evaluating the ability of machine learning models to generalize beyond their training conditions is crucial for practical super‐resolution of fluid flows. In earlier work, we showed that a CNN trained at a single Reynolds number can successfully reconstruct flows at other close Reynolds numbers \citep{LuoBoltzmann2024}. Here, we extend that test to a fundamentally new wake topology, i.e., two identical cylinders placed side-by-side in the spanwise direction following \citet{morimotoGeneralizationTechniquesNeural2022}. The proximity of the cylinders induces complex wake-wake interactions, ranging from gap flows to synchronized vortex shedding, which depend sensitively on their spacing. By applying both CNN and RDN to this two-body configuration, we probe the capacity of each network to infer flow features it has never encountered during training, thus rigorously assessing their robustness and adaptability.

\begin{figure*}
  \centerline{\includegraphics{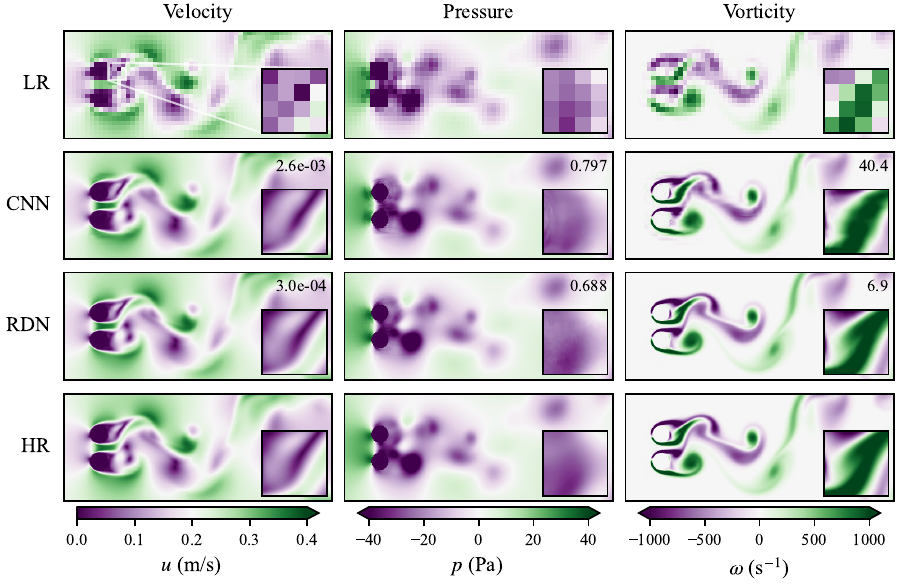}}
  \caption{Low-resolution (LR), super-resolution (CNN and RDN) and high-resolution (HR) flow fields of the flow past two cylinders when $\mathrm{Re}=200$. First column: velocity field, second column: pressure field, third column: vorticity field. The upscaling factor is 8. The distance between the cylinder centers is $1.5d$. The number at the top-right corner indicates the root-mean-square error if applicable. A zoom-in view of a region behind the cylinders is show as an inset at the right-bottom corner.}
  \label{fig:general_macros}
\end{figure*}

Figure~\ref{fig:general_macros} evaluates both networks on an unseen snapshot of flow past two side-by-side cylinders (the center-to-center distance is $1.5d$) at $\mathrm{Re}=200$, using an upscaling factor of 8. The reconstructed velocity, pressure, and vorticity fields reveal that the wake interference produces markedly more intricate patterns than the single-cylinder case. In the velocity field, RDN achieves an error roughly an order of magnitude below that of CNN, yet CNN still captures the overall shear and recirculation with decent fidelity.

The pressure reconstructions expose lingering deficiencies of CNN, i.e., spurious oscillations around each cylinder surface, whereas RDN aligns closely with the high-resolution reference, albeit with slightly elevated noise in the zoomed inset. This suggests that the dense residual pathways of RDN better stabilize steep pressure gradients under complex wake interactions. Vorticity proves to be the most rigorous test. The super-resolved vorticity field from CNN is both blurred and incomplete, with its RMSE exceeding 40. In contrast, RDN delivers fidelity, accurately positioned vorticity lobes with an RMSE of just 6.9, faithfully restoring the alternating eddies that velocity alone cannot reveal.

Efforts to deepen or widen CNN yielded only marginal gains, confirming that its parameter capacity is already fully exploited and that its representational ability has saturated. Only the hierarchical fusion of local and global features of RDN, enabled by dense connections and residual learning, provides the representational power necessary for super-resolving truly complex and multiscale flow fields.

\begin{figure*}
  \centerline{\includegraphics{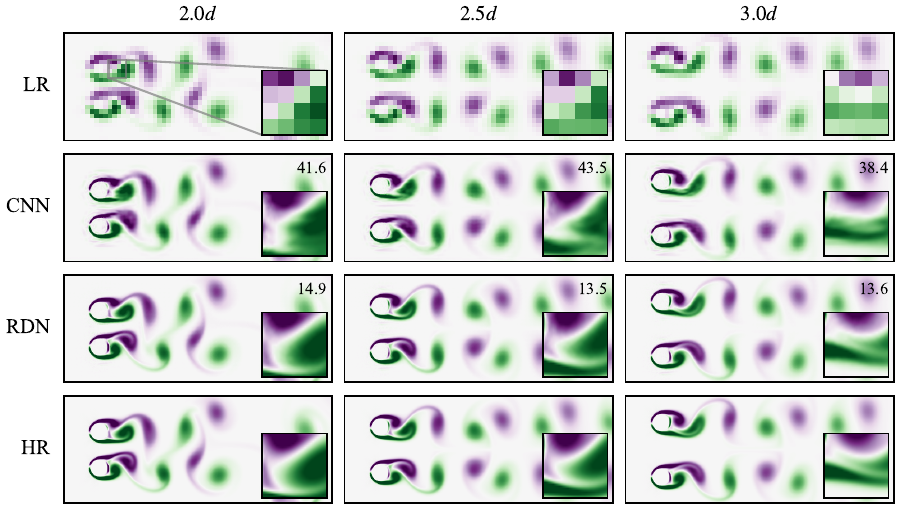}}
  \caption{Low-resolution (LR), super-resolution (CNN and RDN) and high-resolution (HR) vorticity fields of the flow past two cylinders separated by different distances when $\mathrm{Re}=200$. First column: $2.0d$, second column: $2.5d$, third column: $3.0d$. The upscaling factor is 8. The super-resolution models are trained with the data obtained at $1.5d$. The number at the top-right corner indicates the root-mean-square error if applicable. A zoom-in view of a region behind the cylinders is show as an inset at the right-bottom corner. The same colorbar for vorticity in Fig.~\ref{fig:general_macros} has been adopted.}
  \label{fig:general_vordist}
\end{figure*}

To probe true generalization, we challenged both CNN and RDN, each trained only with the $1.5d$-distance case, to reconstruct flows at new separations of $2.0d$, $2.5d$, and $3.0d$ between the cylinder centers. Figure~\ref{fig:general_vordist} shows a representative vorticity field at an extreme upscaling factor of 8. Here, the bicubically downsampled input is so coarse that coherent vortical structures almost vanish. Remarkably, both models infer the correct wake topology, recapturing alternating vorticity cores and shear layers between cylinders, despite never seeing these geometries during training. However, RDN consistently unveils sharper shear‐layer roll‐ups and secondary eddies, attesting to its hierarchical fusion of local and global features. This contrasts with the smoother and less detailed reconstructions from CNN and underscores the superior ability of RDN to extrapolate complex spacing‐dependent flow physics.

\begin{figure*}
  \centerline{\includegraphics{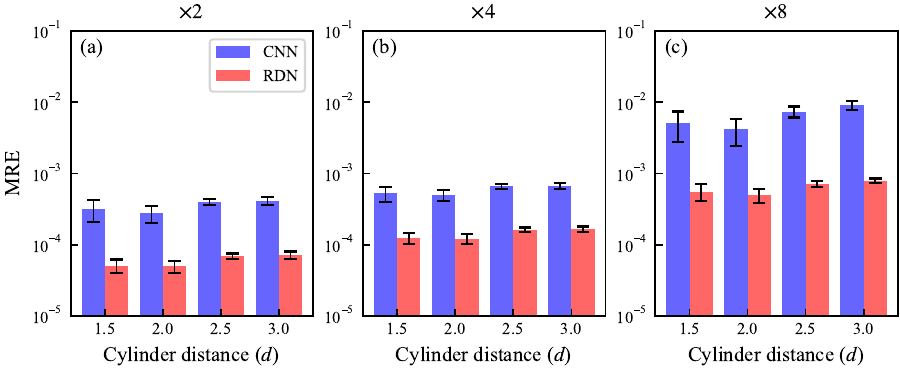}}
  \caption{Comparison of the mean relative errors (MRE) of CNN and RDN regarding the density distribution functions at different upscaling factors: (a) 2, (b) 4, (c) 8. The machine learning models are trained with the data obtained at $1.5d$. The Reynolds number is fixed at 200. The error bars indicate the standard deviations.}
  \label{fig:general_mre}
\end{figure*}

Figure~\ref{fig:general_mre} shows the mean relative error (MRE) of the density distribution functions as the cylinder distance varies from $1.5d$ to $3.0d$ for upscaling factors $r=2$, 4, and 8. Interestingly, both CNN and RDN exhibit MREs that are essentially invariant to cylinder distance, demonstrating that models trained on the most challenging case, tight spacing with strong wake interference, seamlessly generalize to more benign configurations. This robustness contrasts sharply with approaches that reconstruct macroscopic fields directly, where errors can increase by more than an order of magnitude when geometry changes \citep{morimotoGeneralizationTechniquesNeural2022}. We attribute the stability of SRLBM to three key factors: (i) mesoscopic inputs provide richer, transferable physics across scales; (ii) patch-based training and data augmentation, which expose the models to a wide variety of local flow patterns and cylinder positions, see Sec.~\ref{subsec:data}; (iii) fully convolutional architectures impart shift-equivariance, so learned features generalize to cylinders located anywhere in the domain. Thus, even without explicit location conditioning, as in many generative model approaches, our supervised CNN and RDN robustly reconstruct wakes around bluff bodies solely by leveraging mesoscopic distribution learning.

The wider error bars at $1.5d$ and $2.0d$ reflect the higher variability inherent in strongly interacting wakes, whereas the flows at $2.5d$ and $3.0d$ are more periodic and predictable, resulting in tighter uncertainty bounds. Across every spacing and scale, RDN achieves MREs roughly an order of magnitude below those of CNN, translating directly into more faithful super-resolution of macroscopic flow quantities as discussed above.

\section{\label{sec:conclude}Conclusions}

In this work we have introduced SRLBM, a novel multiscale super‐resolution framework that operates directly on the mesoscopic density distribution functions of the lattice Boltzmann method rather than on derived macroscopic variables. By leveraging the more degrees of freedom inherent in discrete populations per lattice site, SRLBM encodes richer physics into a single deep network and then recovers velocity, pressure, vorticity simultaneously. This unified strategy not only streamlines computation but also preserves the natural coupling among flow variables, a feature that is essential for multiphysics simulations requiring concurrent access to diverse flow fields.

Through a systematic comparison of a standard convolutional neural network and a residual dense network, both trained on high‐resolution flow past a cylinder data at Reynolds number of 100, we have demonstrated that RDN delivers superior fidelity across all upscaling factors. The mean relative error of RDN on the distribution functions is an order of magnitude lower than that of CNN once enough training samples are available. When reconstructing macroscopic quantities at upscaling factors up to 8, RDN avoids spurious pressure oscillations and faithfully restores fine‐scale vorticity structures, whereas CNN, despite its adequate performance on velocity, exhibits smoothing and artifacts under severe downsampling.

We have further challenged these models with unseen wake configurations by applying them to flows past two side‐by‐side cylinders at separations of $2.0d$, $2.5d$, and $3.0d$ without retraining (trained using only $1.5d$ data). Both networks recover the correct wake topology even when faced with very coarse input fields ($r=8$), yet RDN consistently resolves sharper shear‐layer roll‐ups and secondary eddies. Moreover, the mean relative error of both models remains essentially invariant to cylinder spacing, in contrast to traditional macroscopic field based methods whose errors escalate by orders of magnitude when geometry changes. This robustness attests to the power of learning based on mesoscopic distributions, which carry transferable multiscale information that facilitates generalization.

Looking forward, the SRLBM paradigm opens several promising avenues, yet this study is limited to low Reynolds number flows, and validating the framework at higher Reynolds regimes will be an important next step. Integrating physical conservation laws and explicit boundary condition constraints directly into the network architecture or loss function could further boost reconstruction accuracy and dramatically reduce the volume of training data required. State-of-the-art super-resolution innovations in computer vision, such as attention-based transformers \citep{xuSuperresolutionReconstructionTurbulent2023} and generative adversarial networks \citep{dengSuperresolutionReconstructionTurbulent2019}, can be adapted to enrich feature representation and sharpen reconstructed flow details. Although our implementation here relies on LBM‐derived numerical snapshots, the ability to recover mesoscopic distribution functions from macroscopic flow measurements means SRLBM can also fuse numerical and experimental data, enabling physics‐guided convergence across various data sources \citep{fukamiObservableaugmentedManifoldLearning2025}. Overall, this study establishes an LBM-based, multiscale framework that offers a compelling alternative for achieving fast and high-fidelity super-resolution reconstructions for fluid dynamics problems.

\begin{acknowledgments}
This research was developed under the support of the national key research and development program of China No.~2020YFA0712500, the Guangdong basic and applied basic research foundation Nos.~2025A1515012962, 2023A1515012881 and 2022B1515120009. The authors thank D.W. David Wang for his insightful discussions on machine learning models for image super-resolution and Xu Han for his support in reviewing this manuscript prior to submission.
\end{acknowledgments}

\section*{Data Availability Statement}

The data that support the findings of this study are available from the corresponding author upon reasonable request.

\nocite{*}
\bibliography{reference}

\end{document}